\documentclass[acp, manuscript]{copernicus}

\begin{document}
\nolinenumbers

\title{Enabling joint chemical and meteorological data assimilation with an atmospheric foundation model}

\Author[1][ejkim1@uw.edu]{Eliot J.}{Kim} 
\Author[1]{Eric J.}{Mei}
\Author[2]{Peter D.}{Ivatt}
\Author[2]{Lee T.}{Murray}
\Author[1][turneraj@uw.edu]{Alexander J.}{Turner}

\affil[1]{Department of Atmospheric and Climate Science, University of Washington, Seattle, WA, USA}
\affil[2]{Department of Earth and Environmental Sciences, University of Rochester, Rochester NY, USA}



\runningtitle{Enabling joint chemical and meteorological data assimilation}

\runningauthor{Kim et al.}

\received{}
\pubdiscuss{} 
\revised{}
\accepted{}
\published{}


\firstpage{1}

\maketitle

\begin{abstract}
The hydroxyl radical (OH) is the main oxidant in the troposphere and accounts for 88\% of the methane sink, yet its abundance and trends remain poorly constrained. Satellites observe many of the chemical and meteorological drivers of OH, but assimilating both requires a model that represents chemistry and meteorology together. Existing chemistry-climate models are too expensive for joint data assimilation. Here, we present AuroraChem, a fine-tuned version of the Aurora atmospheric foundation model that jointly simulates meteorology and the GEOS-Chem chemical fields needed to constrain OH. A one-year AuroraChem rollout takes 12 minutes on one GPU. Free rollouts remain stable for one year but do not fully capture seasonal cycles. Using idealized data assimilation, we find that chemical observations can constrain the meteorological state and meteorological observations can constrain the chemical state. Inserting only the chemical fields raises the annual mean spatial correlation of 500~hPa temperature from 0.89 to 0.92 and recovers the moisture and circulation pattern of the Indian summer monsoon. Inserting only the meteorological fields raises the 500~hPa OH correlation from 0.89 to 0.93, close to the 0.94 from inserting the chemical fields themselves, and recovers the OH seasonal cycle. Inserting only NO$_2$ columns sampled at TROPOMI pixels gives an OH correlation of 0.93. These relationships likely reflect learned covariances rather than physical coupling, and covariances are what data assimilation uses to let an observation of one variable update the others. AuroraChem therefore provides a new pathway to joint chemical and meteorological data assimilation.
\end{abstract}


\introduction  
The hydroxyl radical (OH) is the primary oxidant in the troposphere, oxidizing methane, carbon monoxide, and VOCs to produce ozone and other reactive species. It governs the lifetime of tropospheric pollutants and ozone-depleting substances while also exerting an indirect control on the long-term radiative budget through its regulation of the atmospheric lifetime of gases such as methane \citep[e.g.,][]{lelieveld_global_2016}. As such, OH is arguably the single most important reactive molecule in the troposphere.

Despite its importance, there are large uncertainties in both OH abundance and long-term trends. Global mean OH across 14 chemistry-climate models ranges from 8.7 to 12.8$\times$10$^5$ molec cm$^{-3}$ from 2000--2010 \citep{zhao_inter-model_2019}. Direct OH observations are spatially sparse and have uncertainties of $\pm$30\% \citep[e.g.,][]{stone_tropospheric_2012}. In lieu of direct observations, observations of methyl chloroform (CH$_3$CCl$_3$) have long been used as the main constraint on OH abundance \citep[e.g.,][]{montzka_small_2011,rigby_role_2017,turner_ambiguity_2017} and, as such, the oxidative capacity of the troposphere. Methyl chloroform concentrations have declined to near zero following the Montreal Protocol, which weakens this constraint \citep{liang_deriving_2017}. As a result, observational constraints on OH are weaker today than they were 20 years ago.

Satellites observe many of the physical and chemical drivers of OH. \citet{murray_factors_2014,murray_large_2021} summarize the drivers of global mean tropospheric OH as
\begin{equation}
    \mathrm{[OH]} \propto J_\mathrm{O_3}q\frac{S_N}{S_C^{3/2}}.
    \label{murray}
\end{equation}
where $J_\mathrm{O_3}$ is the photolysis frequency of ozone, $q$ is specific humidity, and $S_N$ and $S_C$ are the sources of reactive nitrogen and reactive carbon, respectively. Ozone photolysis in the presence of water vapor is the primary source of OH. Reactive nitrogen increases OH by recycling peroxy radicals back to OH, and reactive carbon consumes OH. Drivers of global mean OH co-vary with physical meteorology and climate, in which \citet{turner_modulation_2018} and \citet{mei_emulating_2025} showed that OH variability can be explained by the ENSO index. Most of these drivers are well-observed by existing satellites with daily global coverage and have potential to be used for reconstructing spatially-resolved OH fields. For example, satellites observe water vapor (e.g., AIRS), NO$_2$ (e.g., TROPOMI), ozone (e.g., OMI), and CO (e.g., MOPITT) with near-daily global coverage. Prior work has used these observations to estimate OH. \citet{wolfe_mapping_2019} estimated total column OH from satellite HCHO retrievals with a steady-state approximation. \citet{anderson_trends_2024} trained a neural network to predict total column OH from chemistry-climate model fields (including NO$_2$, CO, HCHO, H$_2$O, and O$_3$), then applied it to satellite observations of the same species. However, these satellite-based estimates are limited by the spatiotemporal sparsity of satellite retrievals. Total column OH from \citet{wolfe_mapping_2019} is limited to open oceans where satellite retrievals of HCHO are reliable and available at monthly time resolution due to temporal averaging. Similarly, the total column OH estimates from \citet{anderson_trends_2024} are only available over tropical ocean areas with monthly temporal resolution.

Data assimilation (DA) can combine these observations with a model to produce continuous, global OH fields. \citet{miyazaki_evaluation_2020} assimilated satellite atmospheric composition observations into an ensemble of chemical transport models (CTMs) and reduced the inter-model spread in regional mean OH by 16--50\%. However, CTMs like those used in \citet{miyazaki_evaluation_2020} are driven by offline meteorological inputs, so their chemical state does not feedback to their climate state. Joint assimilation requires a model that represents chemistry and meteorology in a single state, along with the covariances between them. Coupled chemistry-climate models (CCMs) provide this but are computationally expensive. As an example, adding interactive chemistry to CESM2, a state-of-the-art climate model, increases its computational cost by a factor of seven \citep{emmons_chemistry_2020}. Ensemble DA multiplies this cost by the number of ensemble members, typically 10--100 \citep{houtekamer_review_2016}. Prior efforts to assimilate chemistry to constrain meteorology have been limited in scope. \citet{saide_central_2015} assimilated satellite atmospheric composition observations to constrain radiative fields that control boundary layer stability, and \citet{liu_potential_2021} assimilated synthetic satellite NO$_2$ observations to estimate boundary layer winds. However, both studies conducted assimilation on regional domains and covered a time range of days. Joint chemical and meteorological DA with a CCM is therefore computationally prohibitive at global scale and over the multi-year time periods necessary to constrain OH trends.

Machine learning (ML) emulators could make joint DA affordable. ML weather prediction models such as GraphCast, FourCastNet v3, and AIFS match or outperform the ECMWF Integrated Forecast System (IFS) on medium-range forecasts \citep{lam_learning_2023, bonev_fourcastnet_2025, moldovan_aifs_2026}. ML climate emulators run stably for decades to centuries at orders of magnitude lower cost than numerical climate models \citep{cresswell-clay_deep_2025, kochkov_neural_2024, watt-meyer_ace2_2025}. ML emulators of atmospheric chemistry are less mature. \citet{keller_application_2019} replaced the chemical solver in GEOS-Chem with random forests and achieved stable month-long simulations, but at twice the computational cost of the original solver. \citet{kelp_toward_2020} trained a faster neural network emulator of the chemical mechanism, but its simulations became unstable after 1.5 weeks. \citet{mei_emulating_2025} developed a stable coupled chemistry-climate emulator using a linear inverse model, but it simulates column-average quantities and does not capture the nonlinear chemical and physical processes driving OH. Thus, the emulators described above do not provide a joint chemical and meteorological state, stability over the time-scales necessary for DA, and the chemical species needed to constrain OH.

One emulator that may satisfy the above conditions is Aurora \citep{bodnar_foundation_2025}. Aurora is a foundation model pretrained on more than 1 PB of reanalysis, analysis, NWP, and climate model output to learn a general representation of the atmosphere. Fine-tuning adapts the pretrained model to new variables with much smaller training datasets. \citet{bodnar_foundation_2025} fine-tuned Aurora to forecast air quality (ozone, CO, NO$_x$, SO$_2$, and particulate matter) jointly with weather. This is the first ML emulator that jointly simulates chemistry and meteorology on a 3D global grid. However, air quality Aurora does not include OH and the other species needed to constrain oxidative capacity, including Cl and $^{14}$CO. It was also evaluated on 5-day forecasts, not on the longer time-scales necessary for DA over seasons.

Here, we fine-tune Aurora with a parsimonious set of chemical and meteorological fields necessary to constrain OH and the atmospheric oxidative capacity. The fine-tuned model, termed ``AuroraChem" (\url{http://aurorachem.github.io}), is a joint chemistry-meteorology emulator which is accurate for short lead-times and stable over one year. We use an idealized form of DA, ``direct insertion," to test whether chemical observations constrain the meteorological state and whether meteorological observations constrain the chemical state. In our tests, periodic meteorological observations can constrain the chemical state and vice versa. We also show that inserting a single chemical species, even when sampled and column-integrated like a TROPOMI retrieval, constrains OH with similar performance to inserting meteorological fields. These results suggest that AuroraChem could serve as the forward model for joint chemical and meteorological DA.

\section{Fine-tuning an atmospheric foundation model with chemistry}

\subsection{AuroraChem architecture}

\begin{figure}
\noindent\includegraphics[width=\textwidth]{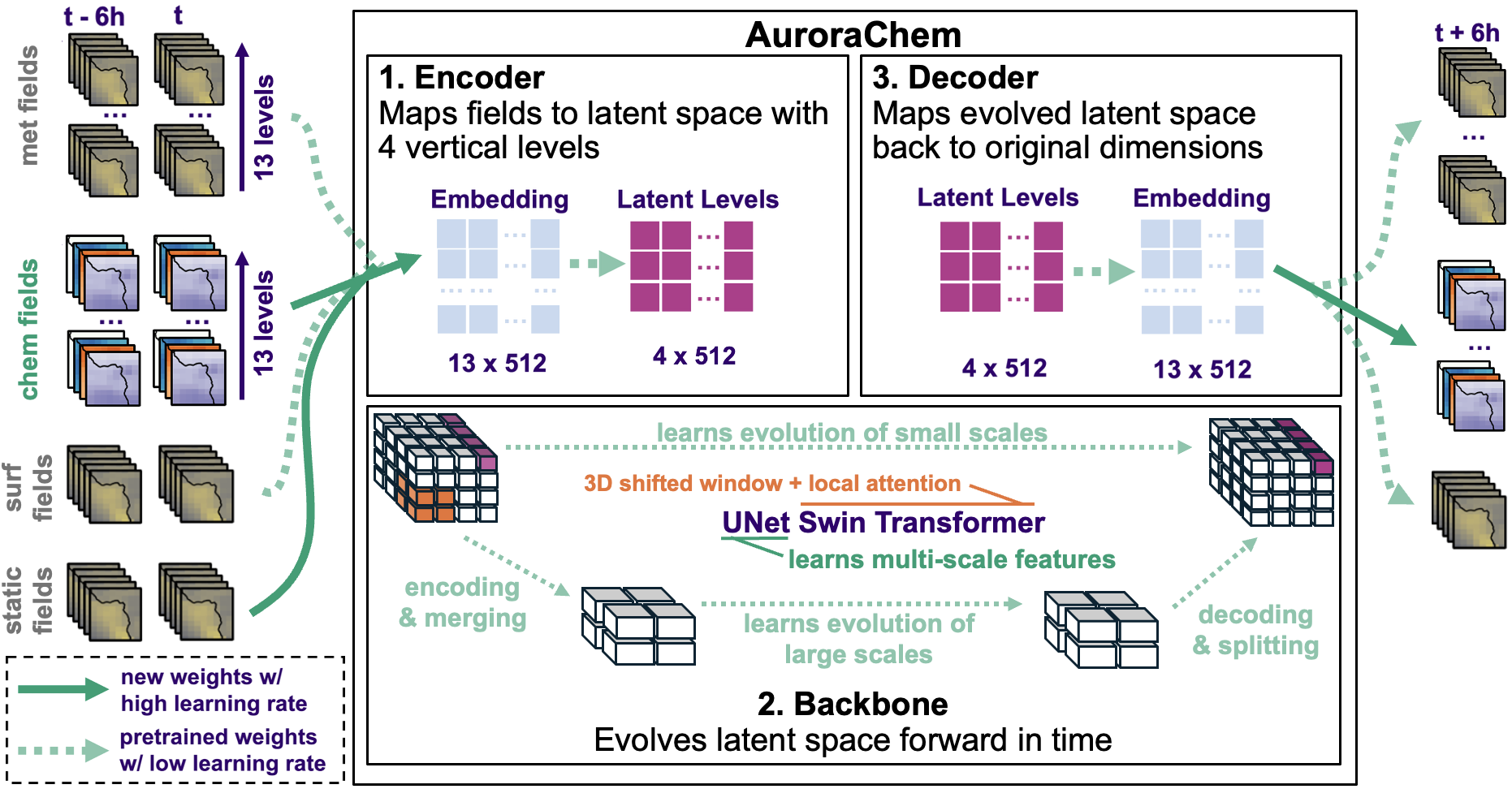}
\caption{\textbf{Schematic of AuroraChem architecture}. The encoder maps meteorological, chemical, surface, and static fields at $t-6$ h and $t$ to a latent state with four vertical levels. The backbone advances the latent state by 6 h. The decoder maps it back to the output fields at $t+6$ h. Solid arrows show new weights trained from scratch at a high learning rate. Dotted arrows show pretrained weights updated at a low learning rate.}
\label{auroraarchitecture}
\end{figure}

\citet{bodnar_foundation_2025} pretrained Aurora to learn a general representation of the atmospheric system. The pretraining dataset included fields from 10 reanalyses, analyses, NWP models, and climate models, totaling over 1 petabyte. \citet{bodnar_foundation_2025} describe the pretraining in detail. The prognostic fields of base Aurora are temperature, specific humidity, zonal and meridional wind, and geopotential height on 13 pressure levels, and 2 m temperature, 10 m zonal and meridional wind, and mean sea level pressure at the surface. Base Aurora also takes static fields as inputs (land-sea mask, soil type, and topography). Aurora ingests the atmospheric states at $t_0 - 6$ h and $t_0$ to forecast the atmospheric state at $t_0 + 6$ h (Figure \ref{auroraarchitecture}). Longer forecasts, or rollouts, apply the model autoregressively, using each forecast as the input for the next step.

Aurora has three components: an encoder, a backbone, and a decoder (Figure \ref{auroraarchitecture}). The encoder first divides the atmospheric fields at each pressure level into patches of size $P \times P \times V \times T$, where $V = 5$ is the number of variables and $T = 2$ is the number of input times. The input patches for each pressure level are embedded into a vector with 512 elements via cross-attention blocks. Crucially, the encoder's learned embedding layer accepts inputs on different grid resolutions and with different sets of variables, which allows pretraining and fine-tuning on diverse datasets. The 13 atmospheric embedding vectors are then compressed with cross-attention into three latent atmospheric levels. Separately, the surface fields are divided into $P \times P \times V \times T$ patches with $V = 4$, embedded, and concatenated to the three latent atmospheric levels. The result is a latent state with dimensions $H \times W \times 4 \times 512$, where $H$ is the number of latitudes divided by $P$ and $W$ is the number of longitudes divided by $P$. 

The latent state is the input to the backbone, which advances it by one $6h$ time step. The backbone's U-Net architecture processes the latent state at several spatial resolutions and represents both small- and large-scale features. At the same time, the backbone propagates information across space and variables via stacked layers of 3D attention blocks. These are Shifted Window (Swin) Transformer blocks, which compute attention within local windows and shift the windows by a fixed offset between layers \citet{liu_swin_2021}. Local attention avoids the quadratic scaling of the cost of global attention with input size, and the shifted windows pass information between neighboring windows. The evolved latent space produced by the backbone is then passed to the decoder. The decoder learns to reverse the encoder's operations to produce the predicted atmospheric and surface fields at $t_0 + 6 h$.

\subsection{Training data}\label{training}

The chemical fields we use for fine-tuning come from a GEOS-Chem simulation for 2018--2019 performed for the FETCH$_4$ project (Fate, Emissions, and Transport of CH$_4$; \url{https://fetch4.github.io/}; last accessed 15 November 2026). GEOS-Chem is a global 3D chemical transport model driven by MERRA-2 meteorological fields that simulates the emissions, transport, chemistry, and deposition of more than 300 trace gases and aerosols \citep[][]{bey_global_2001}. This version (based on GEOS-Chem v14.5.3) adds radiocarbon monoxide ($^{14}$CO) and methane isotopologues to the chemical mechanism. We output 3-hourly, time-averaged GEOS-Chem fields at a 4$^\circ$ by 5$^\circ$ horizontal resolution on 72 vertical levels following hybrid-sigma coordinates.

\begin{table}
\caption{Fields used to fine-tune AuroraChem.}
\label{variables}
\centering
\begin{tabular}{p{4.4cm} p{9.3cm} c c}
\hline
Category (Source) & Names & Levels & Role$^{a}$ \\
\hline
Chemical (GEOS-Chem) & OH, Cl, NO, NO$_2$, O$_3$, CO, $^{14}$CO & 3D & Prognostic \\
Meteorological (ERA5) & T, q, U, V, Z & 3D & Prognostic \\
Surface meteorology (ERA5) & 2m T, 10m U, 10m V, mean sea level pressure & 2D & Prognostic \\
Monthly mean emissions (CAMS) & CO, NO$_x$, NH$_3$, SO$_2$ & 2D & Input \\
Static fields & land-sea mask, soil type, topography, time of day, day of week, day of year & 2D & Input \\
\hline
 \multicolumn{4}{l}{$^{a}$Prognostic fields are both inputs and outputs.}
\end{tabular}
\end{table}

We fine-tune Aurora with seven chemical fields relevant for the atmosphere's oxidative capacity (Table \ref{variables}), and additional chemical fields can be easily added. OH and Cl are the primary chemical sinks of methane in the troposphere, accounting for approximately 88\% and 1\% of the methane sink \citep{saunois_global_2025}. NO and NO$_2$ increase OH through the reaction NO + HO$_2$. Ozone photolysis in the presence of water vapor is the primary source of OH \citep{murray_factors_2014}. CO is the largest single sink of OH \citep{lelieveld_global_2016}. $^{14}$CO is an emerging proxy for OH \citep[e.g.,][]{morgenstern_radiocarbon_2025}. Cosmic rays produce $^{14}$CO mainly at high latitudes and in the upper troposphere and lower stratosphere, and OH is its main sink in the troposphere. $^{14}$CO therefore may provide constraints on the OH seasonal cycle and interhemispheric gradient \citep{petrenko_improved_2021}. 

The meteorological fields come from ERA5 \citep{hersbach_era5_2020}, a state-of-the-art reanalysis produced by the European Center for Medium-Range Weather Forecasts (ECMWF). We use the same fields as base Aurora.  Specifically, we use temperature, specific humidity, zonal and meridional wind, and geopotential height on pressure levels, and 2 m temperature, 10 m zonal and meridional wind, and mean sea level pressure at the surface (Table~\ref{variables}).

In addition to chemistry and climate fields, AuroraChem also ingests emissions, time, and static fields as inputs (Table~\ref{variables}). The emissions are 2019 monthly mean anthropogenic emissions of CO, NO$_x$, SO$_2$, and NH$_3$ from the CAMS global inventory (\url{https://ads.atmosphere.copernicus.eu/datasets/cams-global-emission-inventories}). We use the 2019 emissions for both the training (2018) and test (2019) years. AuroraChem receives no biomass burning or biogenic emissions. The time fields are time of day, day of week, and day of year. The static fields are land-sea mask, soil type, and topography.

We re-grid all fields to a 4$^\circ$ by 5$^\circ$ horizontal resolution and all 3D fields to the 13 isobaric pressure levels present in base Aurora: \{1000, 925, 850, 700, 600, 500, 400, 300, 250, 200, 150, 100, 50\} hPa. The $4^\circ\times{}5^\circ$ resolution reflects the GEOS-Chem simulations available to us. For ERA5, we bilinearly interpolate the pressure-level and surface fields from $0.25^\circ\times{}0.25^\circ$ to $4^\circ\times{}5^\circ$. For GEOS-Chem, we conservatively regrid each model level to the AuroraChem grid, then linearly interpolate each column from the hybrid sigma-pressure levels to the 13 pressure levels. For pressure levels greater than the surface pressure, we use the chemical concentration at the highest pressure level for the "below-ground" pressure levels. We bilinearly interpolate the 2D ERA5, CAMS emissions, and other static fields from their native resolutions to $4^\circ\times{}5^\circ$. We normalize each field at each pressure level before fine-tuning, following \citet{bodnar_foundation_2025}. Meteorological fields are normalized by their mean and standard deviation \citep[Eq.~B5 in][]{bodnar_foundation_2025}. Chemical fields are normalized by half of their time-averaged maximum \citep[Eq.~B8 in][]{bodnar_foundation_2025}.

\subsection{Fine-tuning AuroraChem}

We fine-tune Aurora with GEOS-Chem chemical fields and ERA5 meteorological fields from January 1st through December 31st, 2018. Our fine-tuned model, AuroraChem (\url{http://aurorachem.github.io}), is then evaluated on chemical and meteorological fields that were unseen during training: January 1st through December 31st, 2019. We exclude 30 random samples from the training set to act as the validation set. We fine-tune Aurora for 3,000 steps with a batch size of 1 (60 epochs of 50 batches), about 2 passes through the 2018 data. Fine-tuning takes $\sim$2 h on one NVIDIA H100 GPU. 

To accommodate additional chemical fields and a coarser spatial resolution, we make several changes to the base Aurora architecture. First, we reduce the patch size from $4 \times 4$ to $1 \times 1$. Base Aurora uses $4 \times 4$ patches to reduce cost at 0.25$^\circ$ resolution, but a patch of one grid cell is computationally affordable at $4^\circ\times{}5^\circ$. Second, we train a separate patch embedding for each pressure level, following the Aurora air quality model \citep{bodnar_foundation_2025}. Chemical species vary more with altitude than meteorological fields do. Third, we train a new decoder for the chemical fields and keep the pretrained decoder for the meteorological fields. The separate decoder improves the accuracy of the chemical forecasts.

We update all weights during fine-tuning. The new encoder (7M parameters, 0.5\% of the total) and decoder (18M parameters, 1.4\%) are trained from scratch with a learning rate of $1 \times 10^{-2}$. The pretrained weights, including the backbone (1,240M parameters, 98\%), are updated with a learning rate of $1 \times 10^{-3}$. The loss is a latitude-weighted mean absolute error with a separate weight for each field, following \citet{bodnar_foundation_2025}. The weight for each chemical species is its normalization factor divided by the mean absolute error of a one-step (6 h) persistence forecast \citep[Eq.~D13 of][]{bodnar_foundation_2025}. This gives the forecast error of each field a similar contribution to the loss.

\section{Free-running rollouts of chemistry and meteorology}\label{results}

Our ultimate goal is to conduct data assimilation on seasonal to annual timescales. The base Aurora air quality model was evaluated on 5-day forecasts \citep{bodnar_foundation_2025}, and ML emulators of chemistry have previously been shown to become unstable within weeks \citep{keller_application_2019, kelp_toward_2020}. This raises the question: \emph{``how long can AuroraChem maintain stable and realistic fields?''}

We simulate year-long free-running AuroraChem rollouts. We initialize rollouts with the GEOS-Chem and ERA5 states at 00 and 06 UTC on 1 January 2019, and they receive no further information from the true fields. We note that AuroraChem was fine-tuned with data from 2018: all rollout evaluations are conducted on time stamps unseen during fine-tuning. We compare the rollouts with GEOS-Chem for the chemical fields and with ERA5 for the meteorological fields. A one-year rollout takes about 12 minutes on one NVIDIA H100 GPU. A one-year GEOS-Chem simulation takes about 4,000 minutes on a 32-core Intel Xeon Gold 6226R CPU albeit while simulating roughly 300 chemical species on 72 vertical levels compared to AuroraChem's 7 chemical and 5 meteorological fields on 13 levels.

We evaluate the rollouts with the latitude-weighted spatial correlation coefficient between the predicted field $P$ and the true field $T$ at time $t$:
\begin{equation}\label{rspatial}
    R_s(t) = \frac{\sum_{i, j}w(i)\left(P_{i, j}(t) - \overline{P(t)}\right)\left(T_{i, j}(t) - \overline{T(t)}\right)}{\sqrt{\sum_{i, j}w(i)\left(P_{i, j}(t) - \overline{P(t)}\right)^2\sum_{i, j}w(i)\left(T_{i, j}(t) - \overline{T(t)}\right)^2}},
\end{equation}
\begin{equation}\label{latweight}
    w(i) = \frac{cos(lat(i))}{\frac{1}{H}\sum_{1}^{H}cos(lat(k))}
\end{equation}
where the sums are over the grid cells $(i, j)$ at one pressure level and the overbars denote means over those grid cells. $R_s$ measures whether the spatial patterns of the predicted and true fields match. We weight each grid cell's contribution to $R_s$ by the cosine of its latitude to ensure that the metric is not biased towards performance at the poles. We also use latitude-weighted root mean squared error (RMSE) to evaluate rollouts (Supplemental Figures S3-6): 
\begin{equation}\label{rmse}
    RMSE(t) = \sqrt{\frac{\sum_{i}^{H}\sum_{j}^{W}w(i)(P_{i, j}(t) - T_{i, j}(t))^2}{H * W}}
\end{equation}

\begin{figure}
\noindent\includegraphics[width=\textwidth]{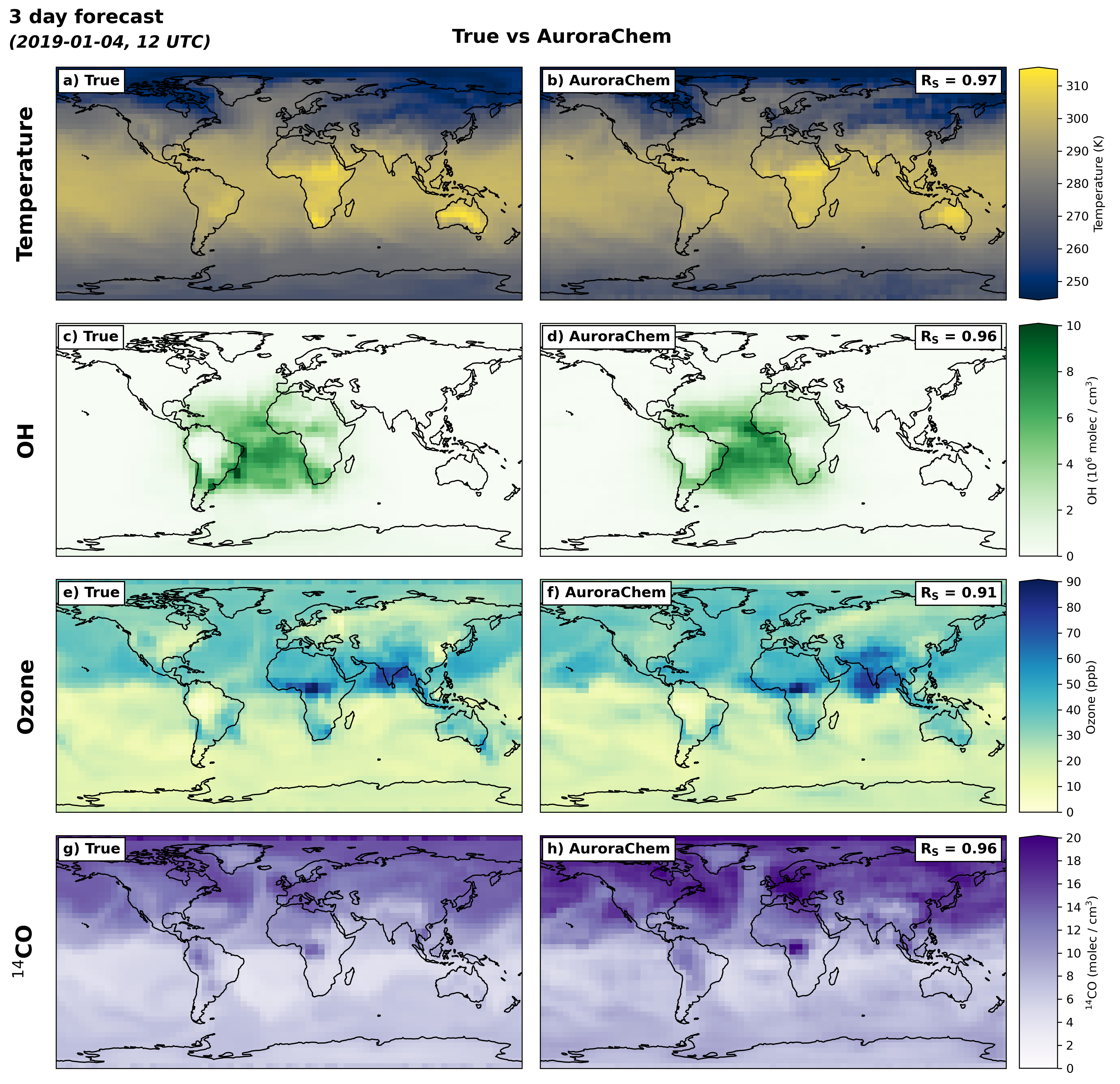}
\caption{\textbf{AuroraChem 3-day forecast at 500 hPa}. Panels show output at 12 UTC on 4 January 2019. AuroraChem was initialized with the states at 00 and 06 UTC on 1 January 2019. The left column shows the true fields: ERA5 temperature (a) and GEOS-Chem OH (c), ozone (e), and $^{14}$CO (g). The right column shows the AuroraChem forecasts (b, d, f, h). The value in each right-hand panel is $R_s$ (Eq.~\ref{rspatial}).}
\label{shortrolloutsamples}
\end{figure}

At short lead times, AuroraChem retains the spatial patterns of both the meteorological and the chemical fields. Figure~\ref{shortrolloutsamples} compares AuroraChem forecasts at 3-day lead times with the true fields. The temperature forecast at 500 hPa has $R_s = 0.97$ and retains the maxima over sub-Saharan Africa and Australia and the minima over Siberia and the Arctic (Figure~\ref{shortrolloutsamples}b). The three chemical species displayed span a range of lifetimes: about 1 s for OH, about 3 weeks for ozone, and about 2 months for $^{14}$CO. The OH forecast at 500 hPa has $R_s = 0.96$ (Figure~\ref{shortrolloutsamples}d). It has OH only in sunlit regions at 12 UTC and low OH over the Amazon, where OH is consumed by biogenic VOCs at the surface. The ozone forecast has $R_s = 0.91$ and retains the interhemispheric gradient and the high ozone over India and sub-Saharan Africa from anthropogenic and biomass burning emissions (Figure~\ref{shortrolloutsamples}f). The $^{14}$CO forecast has $R_s = 0.96$ and retains the interhemispheric gradient and the mid-latitude structure from advection (Figure~\ref{shortrolloutsamples}h). However, it overestimates $^{14}$CO globally. For short lead-times, AuroraChem forecasts maintain the characteristic global and regional spatial patterns for physical fields as well as chemical fields with varying lifetimes.

\begin{figure}
\noindent\includegraphics[width=\textwidth]{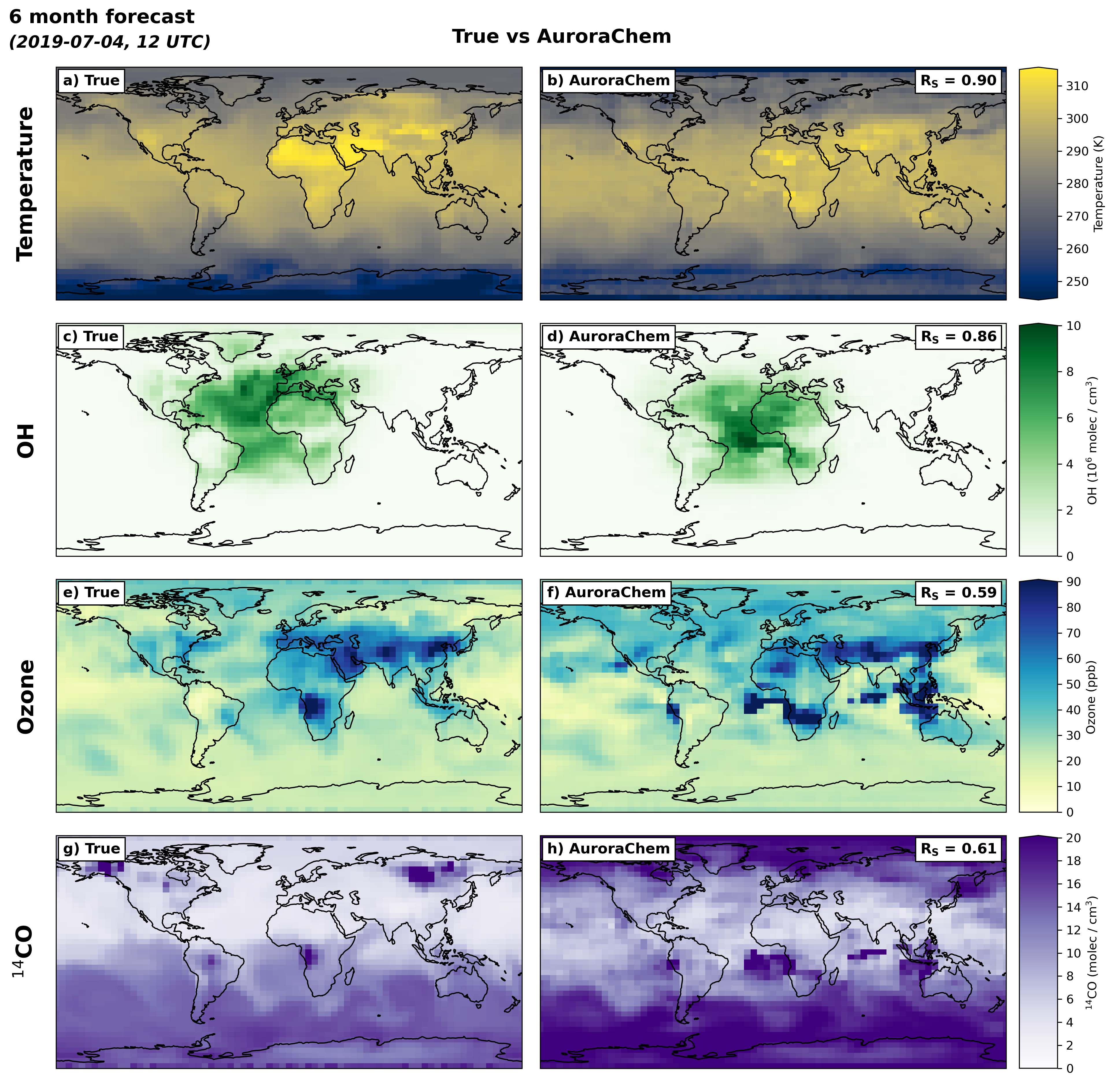}
\caption{Same as Figure~\ref{shortrolloutsamples}, but for the AuroraChem free rollout six months after initialization, valid at 12 UTC on 4 July 2019.}
\label{longrolloutsamples}
\end{figure}

After six months, the AuroraChem rollout remains stable for the physical fields and short-lived species but develops spatial artifacts, particularly for fields without a strong seasonal cycle. Figure~\ref{longrolloutsamples} shows the same fields as Figure~\ref{shortrolloutsamples} but for 4 July 2019 (6 months later). The temperature forecast keeps realistic values and large-scale patterns ($R_s = 0.90$; Figure~\ref{longrolloutsamples}b). The OH forecast from AuroraChem similarly shows realistic concentrations and patterns ($R_s = 0.86$; Figure~\ref{longrolloutsamples}d). It places the OH maximum at the correct longitude for 12 UTC, but it underestimates concentrations in the Northern Hemisphere, especially in the mid-latitudes. AuroraChem therefore gets the position of the OH maximum right for the time of day but does not fully capture the seasonal cycle. We do not observe the development of high-concentration artifacts in the spatial fields of short-lived species.

The longer-lived species develop artifacts, particularly in the tropics. The ozone forecast has $R_s = 0.59$ (Figure~\ref{longrolloutsamples}f). It correctly estimates high ozone concentrations over the Middle East, Northern India, and China. However, it develops spatial artifacts over the Maritime Continent and over the Southern Atlantic. The $^{14}$CO forecast has $R_s = 0.61$ (Figure~\ref{longrolloutsamples}h). It has artifacts in the same tropical regions as ozone and misses the plume over Siberia in GEOS-Chem (Figure~\ref{longrolloutsamples}g). In GEOS-Chem, the $^{14}$CO maximum moves from the Northern to the Southern Hemisphere during Northern Hemisphere summer, when OH is highest in the north. The six-month forecast correctly estimates elevated $^{14}$CO in the Southern Hemisphere but also retains high concentrations in the Northern Hemisphere. Two factors contribute to these errors. First, AuroraChem receives only monthly mean anthropogenic emissions (Section~\ref{training}). In a long-range simulation, it cannot represent transient emission events such as fires without external perturbations to the seasonal cycle, which produce features like the Siberian plume. Second, the 6 h change in a long-lived species is small compared with its spatial variability. The model must learn these small tendencies, and errors at each step accumulate over the rollout. AuroraChem rollouts of the meteorological fields and the short-lived species remain realistic for one year but the longer-lived species drift, and the rollouts do not fully capture the seasonal cycle. This raises the question of whether observations can restore the seasonal cycle and reduce the drift.

\section{Idealized joint chemical and meteorological data assimilation}\label{di}

Joint chemical and meteorological data assimilation requires a forward model in which chemical observations update the meteorological state and meteorological observations update the chemical state. To test whether AuroraChem has learned relationships between chemistry and meteorology that allow this, we conduct idealized data assimilation experiments with direct insertion (DI) \citep{houser_integration_1998}. In DI, the predicted fields are replaced with observed fields every $N$ time steps. We use the true fields as pseudo-observations and insert them every 5 days ($N = 20$ time steps). Each insertion replaces both input states ($t_0 - 6$ h and $t_0$). As such, replacing both the chemical and meteorological fields is equivalent to re-initializing the model. We run three DI configurations for 2019: 1) full DI replaces all 3D fields, 2) met-only DI replaces only the 3D meteorological fields, and 3) chem-only DI replaces only the chemical fields. The 3D meteorological fields are T, q, U, V, and Z while the chemical fields are OH, Cl, NO, NO$_2$, O$_3$, CO, and $^{14}$CO. The free rollout from Section~\ref{results} serves as the control. We evaluate the rollouts with $R_s$ (Eq.~\ref{rspatial}) against GEOS-Chem for the chemical fields and ERA5 for the meteorological fields. All DI experiments are performed on 2019, which is out-of-sample.

\subsection{Constraining global and zonal mean phenomena}

\begin{figure}
\noindent\includegraphics[width=\textwidth]{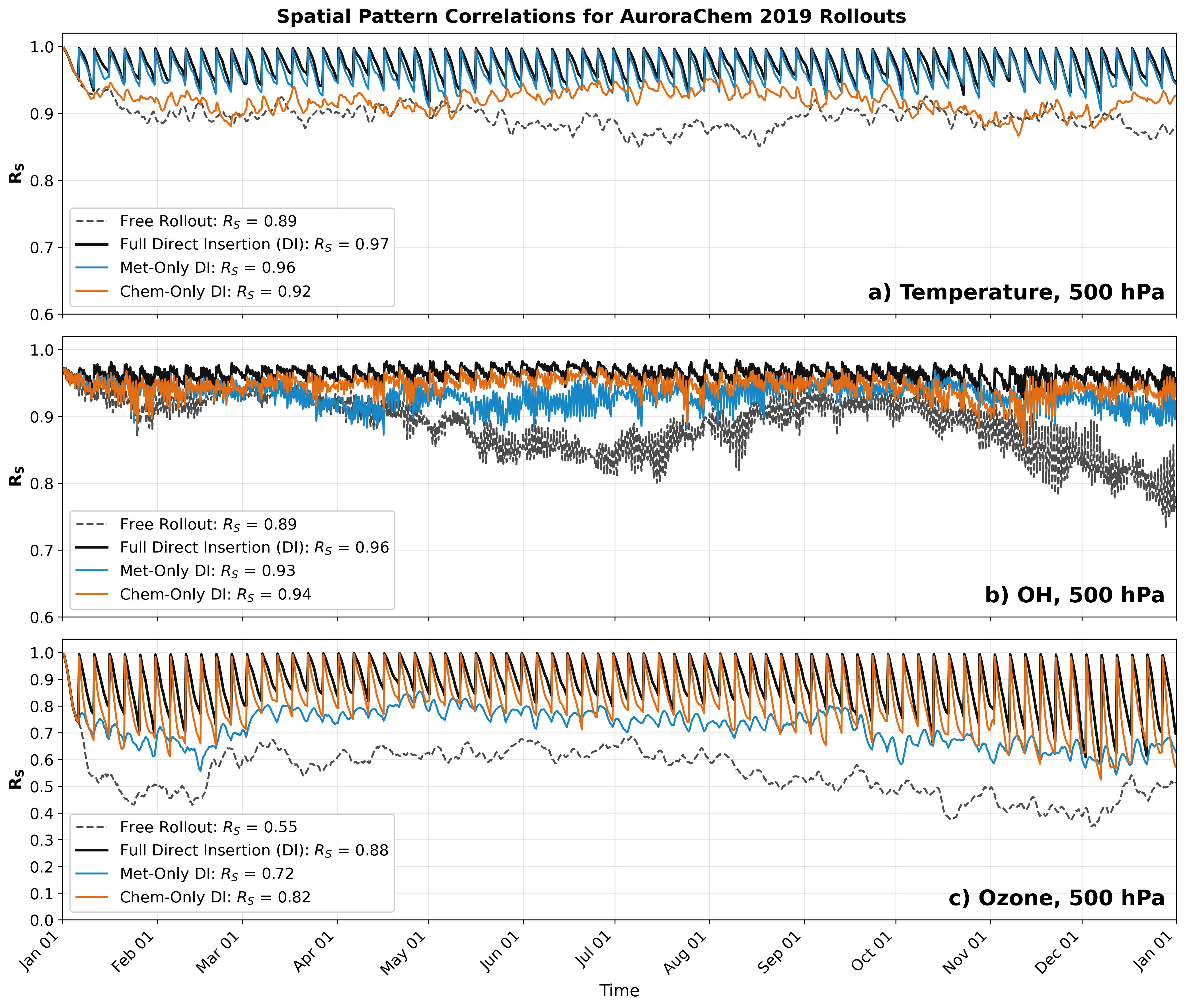}
\caption{Time series of spatial pattern correlations of AuroraChem rollouts compared to the true fields. Panels show $R_s$ (Eq.~\ref{rspatial}) at 500 hPa for (a) temperature, (b) OH, and (c) ozone. Lines show the free rollout (dashed), full DI (black), met-only DI (blue), and chem-only DI (orange). Values in the legends are annual means. Fields are inserted every 5 days, which produces the sawtooth pattern.}
\label{DItimeseries}
\end{figure}

Figure~\ref{DItimeseries}a shows $R_s$ for 500 hPa temperature in each rollout. The free rollout keeps $R_s \approx 0.9$ until May, then falls to $\approx$0.85 in June, which reflects the inability to fully capture the seasonal cycle, seen in Figure~\ref{longrolloutsamples}b. Full DI and met-only DI keep $R_s \geq 0.9$ through the year, with annual means of 0.97 and 0.96, respectively. This is expected, because both insert the true temperature every 5 days. Chem-only DI never inserts temperature, yet it keeps $R_s \approx 0.9$ through the Northern Hemisphere summer. Its annual mean $R_s$ is 0.92, compared with 0.89 for the free rollout. Inserting the chemical fields alone is therefore enough to keep the temperature field in the correct season.

Figure~\ref{DItimeseries}b shows the same comparison for 500 hPa OH. The free rollout keeps $R_s \approx 0.90$ until mid-April, then falls to $\approx$0.80 in mid-June. Full DI and chem-only DI keep $R_s > 0.90$ for most of the year, with annual means of 0.96 and 0.94. This is expected, because both insert the true OH. Similar to the ability of chem-only DI to recover temperature, we find that met-only DI is also able to recover OH with an annual mean $R_s$ of 0.93, close to that of full DI and chem-only DI. OH depends on photolysis and water vapor (Eq.~\ref{murray}), and the meteorological fields carry information about both. As such, inserting only the meteorological fields constrains the OH spatial pattern nearly as well as inserting OH and its chemical drivers. The radiative and dynamical information encoded in meteorological fields is sufficient to constrain the OH spatial pattern.

Ozone rollouts with met-only direct insertion further demonstrate that AuroraChem has learned realistic climate-to-chemistry relationships. Figure~\ref{DItimeseries}c shows the results for 500 hPa ozone. The AuroraChem 500 hPa ozone free rollout develops spatial artifacts at long lead-times (Figure \ref{longrolloutsamples}f); this can be seen in the free rollout decline through the year to $R_s \approx 0.40$ in December. Full DI and chem-only DI keep $R_s \approx 0.80$ until October, with annual means of 0.88 and 0.82. In both, $R_s$ drops from near 1 at each insertion to about 0.7 five days later, indicating that ozone errors grow quickly even from the true state. Met-only DI maintains $R_s$ similar to that of full and chem-only DI five days after each insertion, with an annual mean of 0.72. This demonstrates that AuroraChem has learned sufficient climate-to-chemistry relationships to improve the spatial realism of ozone forecasts using information encoded in meteorological fields, resulting in an ozone field about as accurate as a 5-day forecast from the true ozone.

\begin{figure}
\noindent\includegraphics[width=\textwidth]{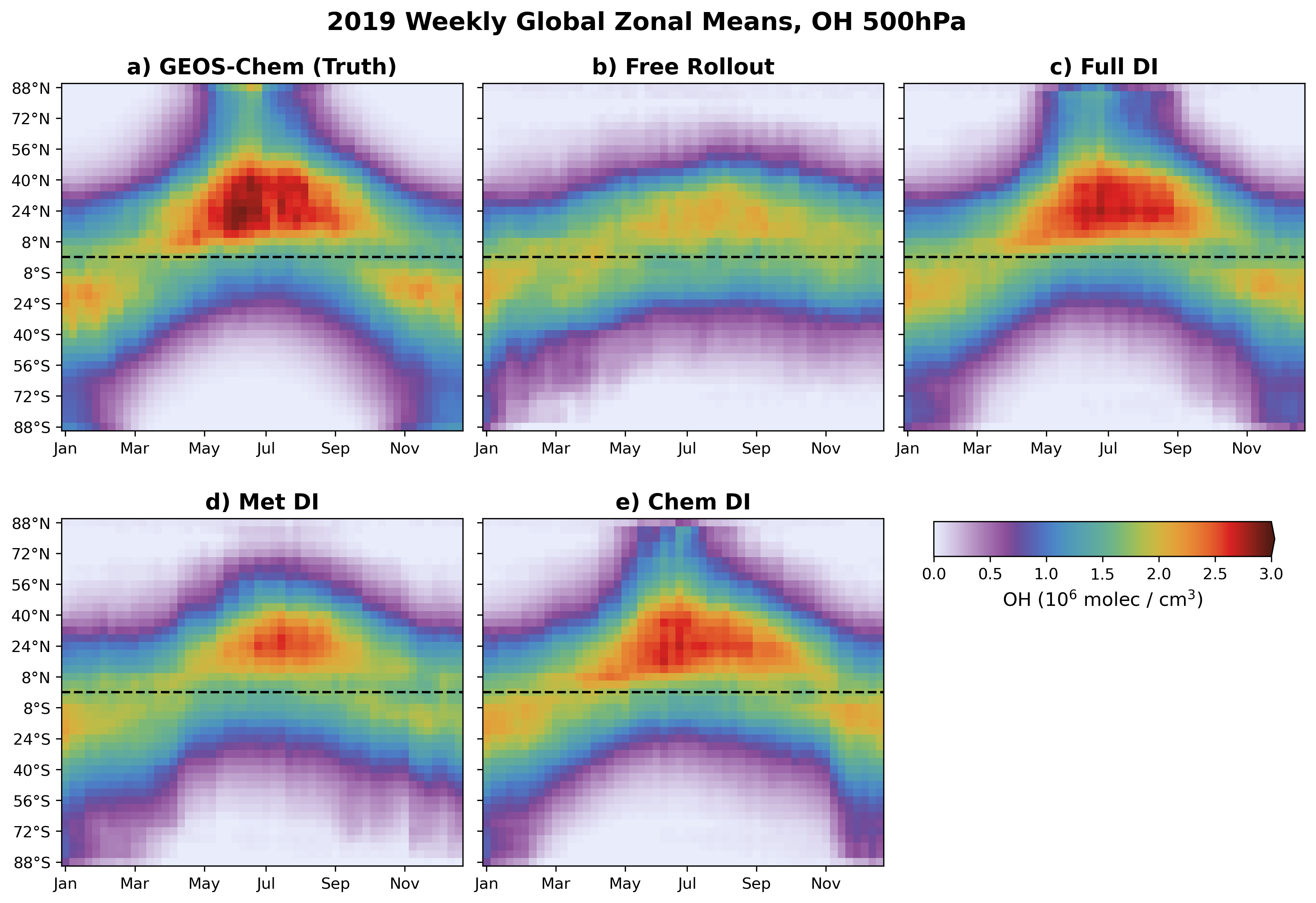}
\caption{Hovm\"{o}ller plots of weekly averaged zonal mean OH at 500 hPa from (a) GEOS-Chem and from AuroraChem rollouts with (b) no DI, (c) full DI, (d) met-only DI, and (e) chem-only DI. The dashed line marks the equator.}
\label{DIhovmollers}
\end{figure}

In Figure \ref{DIhovmollers}, we further demonstrate the ability of each direct insertion configuration to recover the OH seasonal cycle. Figure~\ref{DIhovmollers} shows the weekly zonal mean 500 hPa OH through 2019. In GEOS-Chem, OH peaks in the subtropics of the summer hemisphere and is lowest over the winter pole (Figure~\ref{DIhovmollers}a). The Northern Hemisphere summer maximum exceeds the Southern Hemisphere maximum because of higher ozone and NO$_x$ in the north in GEOS-Chem. The free rollout has a weak OH seasonal cycle and does not represent the magnitude of the subtropical summer maximum (Figure~\ref{DIhovmollers}b). All three DI rollouts better capture the timing and magnitude of the seasonal cycle (Figure~\ref{DIhovmollers}c--e). Met-only DI, which never inserts OH, captures the subtropical summer maxima, the polar winter minima, and the larger Northern Hemisphere maximum, though it underestimates polar summertime OH compared to GEOS-Chem (Figure~\ref{DIhovmollers}d). Inserting only the meteorological fields is able to recover the main spatiotemporal features of the OH seasonal cycle.

\begin{figure}
\noindent\includegraphics[width=\textwidth]{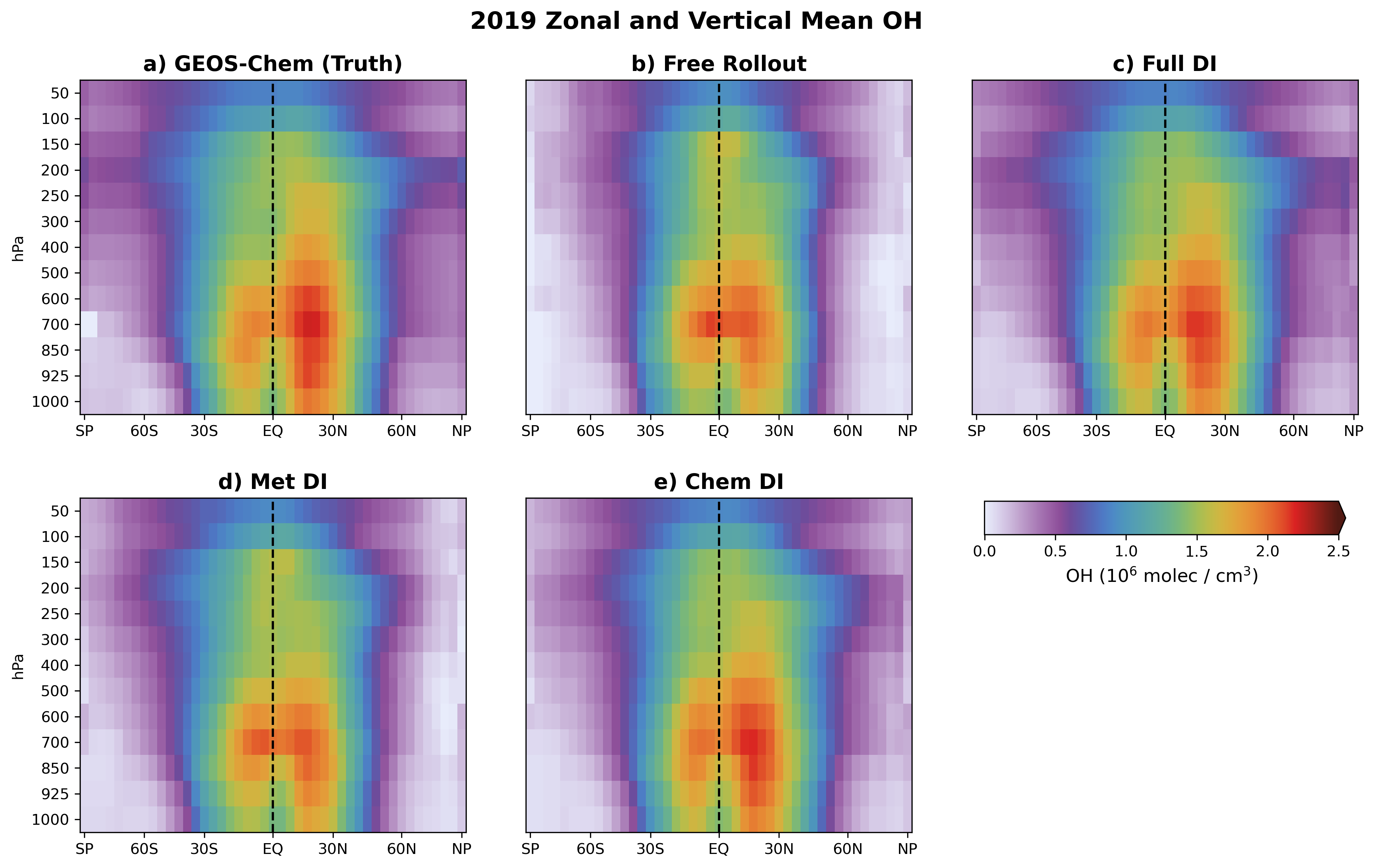}
\caption{Annual zonal mean OH as a function of latitude and pressure from (a) GEOS-Chem and from AuroraChem rollouts with (b) no DI, (c) full DI, (d) met-only DI, and (e) chem-only DI. The dashed line marks the equator.}
\label{globalmeanoh}
\end{figure}

Most of our evaluations thus far have focused on the free troposphere (500 hPa). Figure~\ref{globalmeanoh} shows the 2019 zonal mean OH at all 13 pressure levels. The free rollout captures the general vertical and zonal structure, with higher concentrations in the mid-troposphere over the tropics and lower concentrations near the surface over the equator. However, the free rollout underestimates OH in the northern subtropics and in the upper troposphere over the mid-latitudes (Figure~\ref{globalmeanoh}b). All three DI rollouts better capture the vertical and latitudinal structure of the GEOS-Chem OH (Figure~\ref{globalmeanoh}c--e). The met-only DI rollout estimates greater lower troposphere OH in the Northern Hemisphere than the Southern Hemisphere, better aligning with the GEOS-Chem inter-hemispheric OH gradient (Figure~\ref{globalmeanoh}d). The relationships between chemistry and meteorology that AuroraChem has learned therefore hold throughout the troposphere, not only at 500 hPa.

\subsection{Recovering physical and dynamical features of regional phenomena}

\begin{figure}
\noindent\includegraphics[width=\textwidth]{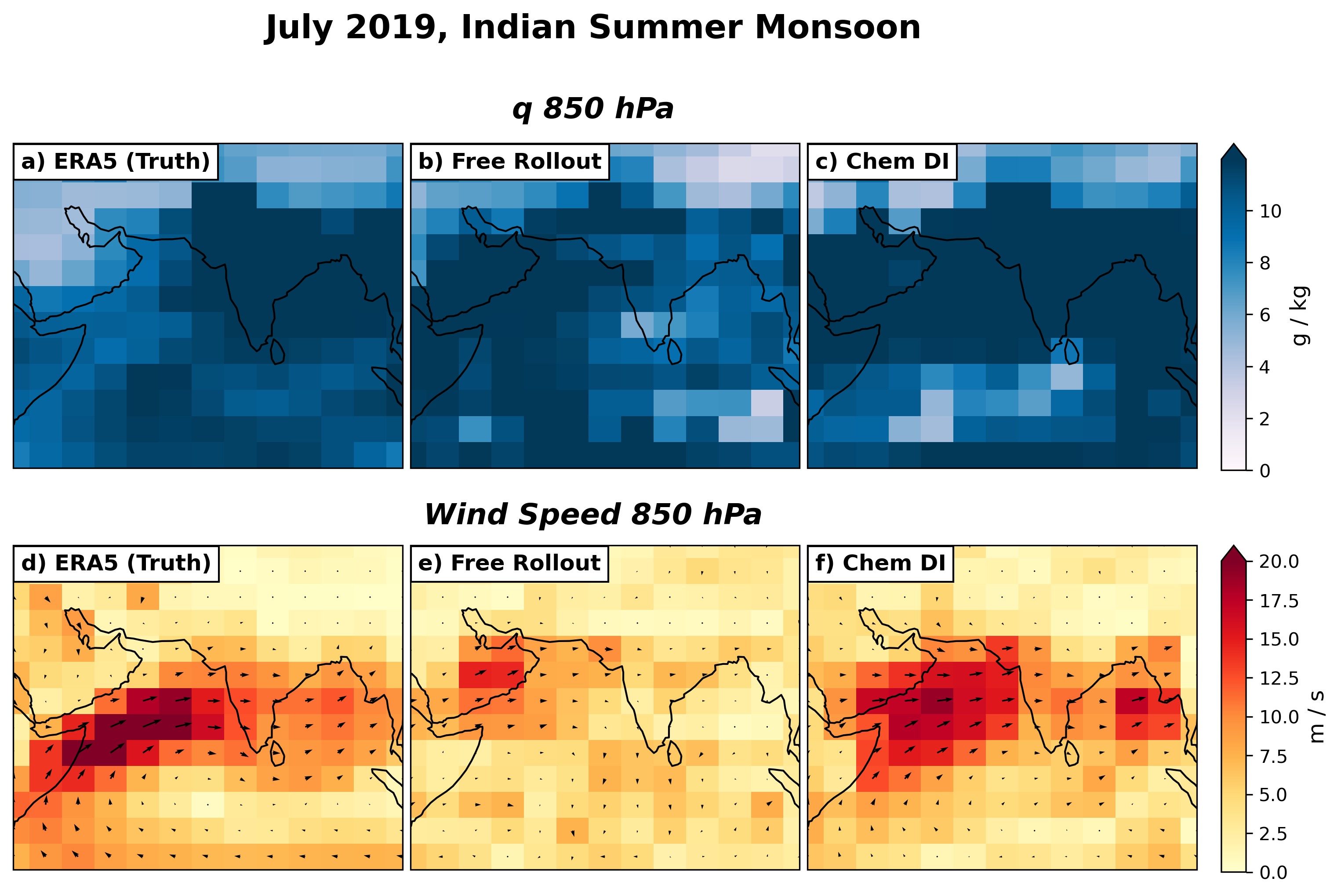}
\caption{July 2019 mean (a--c) specific humidity and (d--f) wind speed at 850 hPa from (a, d) ERA5 and from AuroraChem rollouts with (b, e) no DI and (c, f) chem-only DI. Arrows show the wind direction and magnitude.}
\label{indiansummermonsoon}
\end{figure}

We have demonstrated that meteorological and chemical data can constrain each other at global and zonal mean scales. However, the Indian summer monsoon, a regional circulation that depends on the season, is a more demanding test than global or zonal means. The Indian summer monsoon brings more than 85\% of India's annual rainfall \citep{wang_rainy_2002}, and the free rollout, which stays in Northern Hemisphere winter, does not capture the dynamical characteristics of the monsoon. This raises the question of whether chemical observations alone can recover the Indian summer monsoon. In ERA5, the July mean 850 hPa specific humidity is high over India and Southeast Asia, and a low-level southwesterly jet, the Somali jet, crosses the Arabian Sea from the coast of Somalia to India (Figure~\ref{indiansummermonsoon}a, d). The free rollout places the moisture loading to the west of the ERA5 field, and the southwesterly jet is much weaker and to the north of the ERA5 winds (Figure~\ref{indiansummermonsoon}b, e). Chem-only DI never inserts humidity or wind, yet it recovers both the high humidity over India and the magnitude and position of the southwesterly jet (Figure~\ref{indiansummermonsoon}c, f). However, the chem-only DI rollout overestimates specific humidity over the Arabian Sea, and its jet is slightly weaker and wider than in ERA5. Inserting the chemical fields every 5 days is therefore sufficient to recover the main moisture and circulation features of the monsoon.

\subsection{Constraining AuroraChem with TROPOMI-like NO$_2$ observations}

Thus far, we have assumed full spatial coverage in our hypothetical observing system. However, real observing systems are sparse in space, time, or both. Here we investigate the impact of a more realistic observing strategy on the ability of AuroraChem to recover chemical or meteorological information using an observing system representative of NO$_2$ retrievals from the TROPOspheric Monitoring Instrument (TROPOMI) on Sentinel-5P \citep{veefkind_tropomi_2012}.

TROPOMI retrieves NO$_2$ columns daily (13:30 local time) with near-global coverage, but clouds and retrieval failures leave gaps, and each retrieval measures a column whose vertical sensitivity is described by an averaging kernel. We run three DI experiments, each successively closer to assimilating a real TROPOMI NO$_2$ retrieval to assess the loss of information from each step: 1) NO$_2$-only DI inserts the true 3D NO$_2$ field everywhere, as in Section~\ref{di}, but inserts every time-step and only NO$_2$ rather than all seven chemical species, 2) sampled NO$_2$ DI inserts the true NO$_2$ profile every time-step, but only in grid cells that contain valid TROPOMI pixels within ±3 hours, and 3) sampled NO$_2$ column DI uses the same grid cells as 2) but inserts only the column. The final test uniformly scales the predicted NO$_2$ profile in each cell by the ratio of the true to the predicted NO$_2$ column, with both columns computed using a globally representative TROPOMI NO$_2$ averaging kernel. Therefore, true vertical profile NO$_2$ information is not directly inserted. For the sampled experiments, we use all 2019 reprocessed TROPOMI Level 2 NO$_2$ retrievals with quality assurance values above 0.75, following \citet{sekiya_comparison_2022}, and we interpolate the averaging kernels from the retrieval coordinates to the AuroraChem grid.

\begin{figure}
\noindent\includegraphics[width=\textwidth]{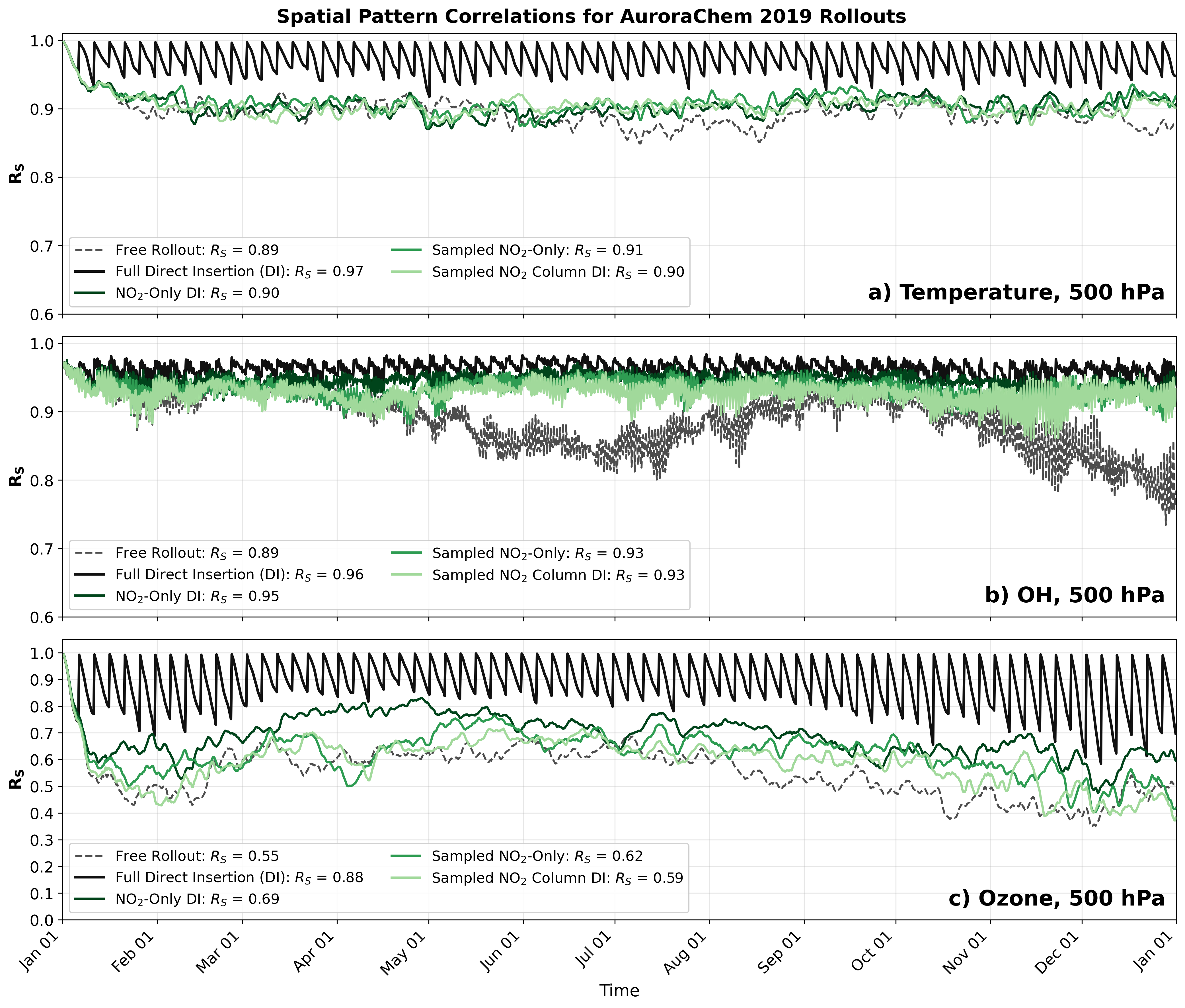}
\caption{$R_s$ (Eq.~\ref{rspatial}) at 500 hPa for the 2019 AuroraChem rollouts with TROPOMI-like NO$_2$ DI for (a) temperature, (b) OH, and (c) ozone. Lines show the free rollout (dashed), full DI (black), NO$_2$-only DI (dark green), sampled NO$_2$ DI (medium green), and sampled NO$_2$ column DI (light green). Values in the legends are annual means.}
\label{tropDItimeseries}
\end{figure}

Figure~\ref{tropDItimeseries}a shows $R_s$ for 500 hPa temperature in the NO$_2$ DI experiments. All three NO$_2$ DI experiments have an annual mean temperature $R_s = 0.90$ to $0.91$, comparable to the free rollout $R_s$ of 0.89. NO$_2$ information alone is not sufficient to recover the seasonal cycle of temperature. Figure~\ref{tropDItimeseries}b shows the 500 hPa OH $R_s$ of NO$_2$ DI experiments. NO$_2$-only DI keeps OH $R_s \geq 0.90$ through 2019, with an annual mean of 0.95, and the sampled and column experiments both have annual means of 0.93, compared with 0.89 for the free rollout. Daily, sparse NO$_2$ column information constrains OH nearly as well as complete 3D NO$_2$ fields.

Figure~\ref{tropDItimeseries}c shows the NO$_2$ DI results for 500 hPa ozone. NO$_2$-only DI raises the annual mean $R_s$ from 0.55 in the free rollout to 0.69, comparable to the $R_s$ value full DI reaches five days after each insertion. Sampling at TROPOMI pixels lowers the annual mean $R_s$ to 0.62, even with daily insertion, and inserting only column information lowers $R_s$ to 0.59, still above the free rollout. Ozone is more sensitive than temperature or OH to how NO$_2$ is inserted. Sparse NO$_2$ column observations therefore constrain OH, partly constrain ozone, and do not constrain temperature in AuroraChem rollouts. It is important to caveat that these experiments use noise-free pseudo-observations, so they test the effects of sparse sampling and column integration but not the effect of retrieval errors.

\subsection{What is AuroraChem learning?}

We have shown that inserting the chemical fields constrains the meteorological state in AuroraChem and inserting the meteorological fields constrains the chemical state. This raises the question \emph{``Is AuroraChem learning the physics that links chemistry and meteorology, or just the correlations between them?''} We posit here that AuroraChem is primarily learning the correlation structure, but that this is still sufficient for the purposes of data assimilation. Some of the relationships we find are consistent with the underlying physics. For example, met-only DI recovers the OH seasonal cycle, as we would expect from the dependence of OH on photolysis and water vapor (Eq.~\ref{murray}). However, other relationships cannot be physical. Chem-only DI recovers the moisture and the Somali jet of the Indian summer monsoon, yet chemistry does not drive the monsoon on 5-day time scales, and the training data contain no feedback from chemistry to meteorology because GEOS-Chem is driven by prescribed meteorology. Instead, the chemical fields carry an imprint of the meteorology through transport, photolysis, and the seasonal cycle, and AuroraChem uses this imprint to infer the meteorological state. As such, it is likely that AuroraChem has learned the correlation structure between the chemical and meteorological fields, not necessarily the physics that produces it. 

Consider the analogous problem of keeping a car on the road through a turn. There is a lot of physics behind this that depends on the conditions (e.g., speed, tire friction, ice on the road), but drivers do not solve these equations. Instead, they learn from experience which steering input corresponds to which view of the road. These learned correlations work well on roads similar to those the driver has seen before but fail on ice. We expect AuroraChem to behave similarly when perturbed outside of its training data. For example, AuroraChem was trained with the monthly mean emissions for one year, so it has never seen how chemistry responds to inter-annual changes in emissions, and it is unlikely to produce the correct OH response to a perturbation in NO$_x$ emissions. This means AuroraChem is not well suited for perturbation experiments that attribute changes in OH to their underlying drivers. Nevertheless, AuroraChem is well suited to some data assimilation applications. Data assimilation requires knowing how an observation of one variable should update the others, and this is determined by the correlation structure. The goal of joint chemical and meteorological data assimilation is to keep OH consistent with the observations, much like the goal of the driver is to keep the car on the road, and the correlations learned by AuroraChem are sufficient for this purpose.

\section{Conclusions}

Constraining OH requires a model that can assimilate both chemical and meteorological observations. Coupled chemistry-climate models are presently too computationally expensive to use for this task. In this work, we present AuroraChem (\url{http://aurorachem.github.io}), a fine-tuned version of the Aurora atmospheric foundation model with the chemical fields needed to constrain OH. Fine-tuning Aurora on one year of chemical and meteorological fields produces a model (AuroraChem) that simulates a one-year rollout in about 12 minutes on one GPU. Free rollouts from AuroraChem remain stable for one year but do not fully capture seasonal cycles, and the longer-lived species develop artifacts in the tropics. Using an idealized data assimilation framework, we find that assimilating chemical fields allows the model to recover the seasonal cycle of the meteorology, and assimilating meteorological fields recovers the seasonal cycle of OH. Assimilating only NO$_2$, sampled daily like TROPOMI, constrains OH nearly as well as inserting complete NO$_2$ fields. These results show that a fine-tuned atmospheric foundation model provides a new pathway to joint chemical and meteorological data assimilation.

AuroraChem builds on a pretrained foundation model rather than an emulator trained from scratch. We fine-tune Aurora with seven chemical species from a GEOS-Chem simulation (OH, Cl, NO, NO$_2$, ozone, CO, and $^{14}$CO) and five meteorological fields from ERA5, regridded to $4^\circ\times{}5^\circ$ and 13 pressure levels. Fine-tuning requires three changes to the architecture: a patch size of one grid cell, a separate patch embedding for each pressure level, and a separate decoder for the chemical fields. The pretrained backbone holds 98\% of the 1.3 billion parameters and is updated at a low learning rate, so AuroraChem retains most of the pretrained model. A one-year AuroraChem rollout takes about 12 minutes on one H100 GPU, compared with about 4,000 minutes for a one-year GEOS-Chem simulation on a 32-core CPU.

We evaluate AuroraChem with free rollouts initialized on 1 January 2019. After three days, the 500 hPa forecasts retain the spatial patterns of temperature, OH, ozone, and $^{14}$CO, with $R_s$ between 0.91 and 0.97. After six months, temperature and OH keep realistic values and spatial patterns ($R_s$ of 0.90 and 0.86, respectively), but the rollout state retains some features of Northern Hemisphere winter. OH peaks at the correct longitude for the time of day but does not fully capture the northward shift in the maxima during Northern Hemisphere summertime. The longer-lived species drift further, with $R_s$ of 0.59 for ozone and $0.61$ for $^{14}$CO, and develop artifacts over the Amazon, the Congo basin, and the Maritime Continent. These errors likely reflect the single year of training data, the absence of transient emissions such as fires, and the small 6 h tendencies of long-lived species.

We use direct insertion to test whether observations of chemistry constrain meteorology, and vice versa. Inserting only the chemical fields every 5 days keeps 500 hPa temperature at an annual mean $R_s$ of 0.92, compared with 0.89 for the free rollout, and recovers the moisture loading and the southwesterly Somali jet of the Indian summer monsoon. Inserting only the meteorological fields keeps 500 hPa OH at 0.93, compared with 0.89 for the free rollout and 0.94 for inserting the chemical fields themselves. Inserting meteorology also recovers the latitudinal and vertical structure of the OH seasonal cycle and raises ozone $R_s$ from 0.55 to 0.72. These relationships are statistical. AuroraChem has likely learned covariances between the chemical and meteorological fields rather than the physical coupling between them. However, these covariances are what data assimilation uses to let an observation of one variable update the others.

We then test whether sparse, column-integrated observations of a single species can constrain AuroraChem. Inserting only NO$_2$ at every time-step keeps OH at an annual mean $R_s$ of 0.95. Inserting NO$_2$ 1) only in grid cells with valid TROPOMI pixels and 2) only as columns computed with a globally represntative TROPOMI averaging kernel, gives annual mean OH $R_s$ of 0.93 for both. Ozone is more sensitive to the NO$_2$ sampling. Its $R_s$ falls from 0.69 with complete NO$_2$ fields to 0.59 with TROPOMI-like columns, though it stays above the free rollout (0.55). Sparse NO$_2$ column observations can therefore constrain OH in AuroraChem.

AuroraChem could serve as the forward model for joint assimilation of chemical and meteorological observations to constrain OH. Its low cost makes ensemble data assimilation practical. Its learned covariances can inform the cross-variable terms of the background error covariance, which are difficult to specify by hand and have limited the application of data assimilation to chemistry-climate models \citep[e.g.,][]{bocquet_data_2015}. Because it is differentiable, it provides the gradients for 4D variational data assimilation without a hand-coded adjoint model \citep[e.g.,][]{vonich_predictability_2024}. Several areas for improvement remain. Fine-tuning on multiple years of chemistry-climate model output should improve the seasonal cycle and add the feedbacks from chemistry to meteorology that GEOS-Chem lacks. Fine-tuning at $1^\circ\times{}1^\circ$ would better resolve short-lived species such as OH, and adding CH$_4$ and its isotopologues ($^{13}$CH$_4$, $^{12}$CH$_3$D, and $^{13}$CH$_3$D) would allow assimilation of methane observations. The next step is to assimilate real meteorological and chemical observations, including TROPOMI NO$_2$, to estimate three-dimensional, time-resolved OH. Such estimates would constrain recent changes in OH, and its role in the methane budget, independently of methyl chloroform.


\codeavailability{The code for this study is available at https://github.com/aurorachem/AuroraChem (last access: 16 November 2026) and code will be uploaded to Zenodo upon acceptance.}


\authorcontribution{E.J.K. and A.J.T. designed research; E.J.K. and A.J.T. performed research; E.J.K., E.J.M., P.D.I., L.T.M., and A.J.T. analyzed data; and E.J.K., E.J.M, P.D.I., L.T.M., and A.J.T. wrote the paper.} 

\competinginterests{The authors do not declare any competing interests.} 

\begin{acknowledgements}
This work is part of the FETCH$_4$ project, and support was provided by Schmidt Sciences. E.J.M. acknowledges funding from a National Defense Science and Engineering Graduate Fellowship from the Department of Defense, supported by the Air Force Office of Scientific Research under award number FA9550-25-C-B010 in the dollar amount of \$143,000.
\end{acknowledgements}


\bibliographystyle{copernicus}
\bibliography{bibliography.bib}

\end{document}